\ifdefined\pdfoutput\pdfoutput=1\fi
\documentclass[conference,compsoc]{IEEEtran}
\IEEEoverridecommandlockouts

\usepackage{iftex}
\ifXeTeX
  \usepackage{fontspec}
\fi
\usepackage{cite}
\usepackage{amsmath,amssymb,amsfonts}
\usepackage{graphicx}
\usepackage{booktabs}
\usepackage{multirow}
\usepackage{xcolor}
\usepackage{url}
\usepackage{tikz}
\usetikzlibrary{arrows.meta,positioning,shapes.geometric}

\usepackage{eso-pic}

\AddToShipoutPictureFG*{%
  \AtPageUpperLeft{\put(\LenToUnit{0.5\paperwidth},\LenToUnit{-0.45in}){%
    \makebox[0pt][c]{\footnotesize\itshape Accepted to the Workshop on AI for Cyber Threat
    Intelligence (WAITI 2026), co-located with the Annual Computer Security Applications
    Conference (ACSAC 2026).}}}%
  \AtPageLowerLeft{\put(\LenToUnit{0.5\paperwidth},\LenToUnit{0.3in}){%
    \makebox[0pt][c]{\parbox{0.85\paperwidth}{\centering\scriptsize
    \copyright\,2026 IEEE. Personal use of this material is permitted. Permission from IEEE
    must be obtained for all other uses, in any current or future media, including
    reprinting/republishing this material for advertising or promotional purposes, creating new
    collective works, for resale or redistribution to servers or lists, or reuse of any
    copyrighted component of this work in other works.}}}}%
}

\newcommand{\cav}{\textsc{CAV}}
\newcommand{\abstain}{\textsc{Abstain}}

\begin{document}

\title{When AI Investigators Cannot Trust Their Tools:\\
Conflict-Aware Evidence Verification for Cyber Threat Investigation}

\author{\IEEEauthorblockN{Yuchen Zhang}
\IEEEauthorblockA{\textit{Khoury College of Computer Sciences, Northeastern University}\\
Seattle, WA, USA\\
zhang.y19@northeastern.edu}}

\maketitle

\begin{abstract}
LLM agents increasingly triage security alerts by combining the outputs of several detection
tools, implicitly treating those tools as trustworthy. Unlike prior work on conflicts between a
model's knowledge and its context, we study conflicts \emph{among the tools themselves}: what
happens when this machine-generated evidence is conflicting or wrong. On the corrected
CIC-IDS2017 dataset, four heterogeneous detectors emit labels and confidences that we corrupt
under a threat model varying the number of corrupted tools, the persistence of the corruption,
and the confidence attached to wrong outputs. Across open-weight LLMs from 3B to 14B parameters,
unverified investigators are fragile: one corrupted detector out of four lowers naive accuracy to
.50--.72, three lower every unverified investigator to .13 or less, and honest disagreement alone
makes them abstain on up to 63\% of flows. We propose Conflict-Aware Evidence Verification
(\cav{}), built from established components: an equal-weight vote that ignores reported
confidences, an evidence-conflict gate, and an independent nearest-neighbour trust check that
tests the leading candidates. With its prompt frozen before testing, \cav{}-LLM improves on the
naive investigator in every corrupted condition ($p < 0.001$), reaching .84--.88 with three
corrupted detectors against at most .08, while calling the LLM on 0.4\% of clean traffic. A fixed
verification rule is as accurate or better unless most tools are corrupted. When we make the
independent channel itself noisy, \cav{} inherits its errors: it stays ahead when two or more tools
are corrupted, but adding the channel as an extra voter is better when at most one is. Verification
thus moves trust from the tools to the check, which must be measured and protected in turn.
\end{abstract}

\begin{IEEEkeywords}
LLM agents, cyber threat intelligence, intrusion detection, evidence conflict, robustness,
selective prediction
\end{IEEEkeywords}

\section{Introduction}
\label{sec:intro}

Security operations centers (SOCs) face alert volumes in which most alarms are false
positives~\cite{alahmadi2022false}, and LLM agents are now being placed in the loop to triage
alerts and draft assessments~\cite{wei2025cortex,wu2026excytin}. Such an agent rarely inspects
raw traffic. It reasons over \emph{machine-generated evidence}: the labels and confidences
emitted by intrusion detection systems, classifiers, and anomaly detectors.

This design assumes that the tools are trustworthy. In practice, detectors disagree, degrade
under drift~\cite{arp2022dos}, and can be misconfigured, poisoned, or compromised. Because their
outputs enter the agent's context, they are also an attack surface~\cite{greshake2023indirect}.
LLMs are known to adopt contradicting context over their own knowledge~\cite{xie2024chameleon}
and to defer to confident assertions~\cite{sharma2024sycophancy}. We therefore ask: \emph{when
an AI investigator receives conflicting or unreliable machine-generated evidence, does it
identify the unreliable evidence, or propagate it into the final threat assessment?}

Existing work frames evidence conflict for LLMs as a conflict between what a model knows and
what it reads~\cite{xie2024chameleon,wu2024clasheval}, or between retrieved documents of unknown
reliability~\cite{xu2024conflicts}. Tool-mediated investigation is a different regime, which we
call \emph{inter-tool conflict}. The agent has essentially no prior about an individual flow,
so its verdict is decided entirely by how it reconciles the tools; each source is a predictor with
a measurable error rate, so a principled answer exists (Dawid--Skene estimation~\cite{dawid1979maximum},
weak supervision~\cite{ratner2017snorkel}, truth discovery~\cite{li2016truth}); and the confidence
attached to each claim is itself an attack surface.
LLMs over-trust a \emph{single} faulty tool~\cite{sun2024toolsfail}, but to our knowledge no prior
work measures how they resolve conflicts \emph{among several} corrupted tools, or compares them
with the statistical aggregators built for this problem.

We present a controlled study of this behaviour and a lightweight mitigation; we do not propose
a new detector. We instrument the \emph{evidence layer} between four heterogeneous detectors and
the investigating agent on the corrected CIC-IDS2017 dataset~\cite{sharafaldin2018cicids,
liu2022error}, corrupt a controlled share of the evidence, and ask how robust LLM investigators
are (RQ1), how much the confidence a tool reports influences them (RQ2), whether conflict-aware
verification helps, also when the independent evidence is itself noisy (RQ3), and at what cost
(RQ4). Our contributions are:
\begin{enumerate}
  \item a threat model and testbed for \emph{inter-tool} evidence conflict in AI-assisted threat
        investigation that separates the number, persistence, and confidence of corrupted
        tool outputs (Sections~\ref{sec:threat}--\ref{sec:testbed});
  \item a controlled evaluation of open-weight LLM investigators (3B--14B) against statistical
        aggregators, showing that they over-escalate honest disagreement, follow corrupted
        majorities, and are paralysed rather than persuaded by confidently wrong tools;
  \item Conflict-Aware Evidence Verification (\cav{}), built entirely from established
        components, which improves accuracy in every corrupted condition while calling the LLM
        on 0.4\% of clean traffic, with an ablation showing that a leaner variant is stronger, a
        failure analysis tracing the remaining errors to \emph{misread} evidence, and a first test
        with a degraded independent channel (Sections~\ref{sec:method}--\ref{sec:analysis}).
\end{enumerate}

\section{Related Work}
\label{sec:related}

\paragraph{LLMs for threat analysis}
Benchmarks evaluate LLM knowledge and reasoning on threat intelligence~\cite{alam2024ctibench}
and agents that investigate incidents over security logs~\cite{wu2026excytin}. Multi-agent
systems triage SOC alerts~\cite{wei2025cortex}, and agents also perform offensive tasks such as
penetration testing~\cite{deng2024pentestgpt}, and content-grounded agents answer from supplied
literature~\cite{ma2026intragent}. These works typically assume that their inputs are correct.

\paragraph{Untrusted context}
Indirect prompt injection hijacks tool-using agents through adversarial
instructions~\cite{greshake2023indirect,debenedetti2024agentdojo}. Our threat model is quieter:
corrupted tools inject no instructions, only plausible but wrong \emph{data}, so instruction-level
defences do not apply; memory-based attacks likewise evade such detectors~\cite{rao2026fragfuse}. LLMs often accept silently wrong outputs of a single faulty
tool~\cite{sun2024toolsfail}. We study several tools that contradict one another. Research on knowledge conflicts distinguishes context--memory, inter-context, and
intra-memory conflicts~\cite{xu2024conflicts}. LLMs readily adopt incorrect retrieved context,
especially when their prior is weak~\cite{xie2024chameleon,wu2024clasheval}, and defer to
confident assertions~\cite{sharma2024sycophancy}. Unlike free-text documents, our sources are predictors with
measurable error rates and an attackable confidence field.

\paragraph{Combining and verifying unreliable sources}
Dawid--Skene~\cite{dawid1979maximum}, weak supervision~\cite{ratner2017snorkel}, and truth
discovery~\cite{li2016truth} estimate per-source error rates from agreement, and log-odds weights
are optimal for independent sources of known accuracy~\cite{nitzan1982optimal}.
Score-level fusion, e.g., for fraud detection~\cite{han2026cfa}, assumes that every member reports
honestly, and aggregation of multiple LLMs exploits their correlation~\cite{ai2026beyond}. We study
the reverse configuration, in which an LLM aggregates detectors whose corrupted errors are
maximally correlated.
Whether to trust a prediction can be checked against nearby training data~\cite{jiang2018trust},
and selective classification trades coverage for risk~\cite{geifman2017selective}. We place an
LLM in the aggregator's position, measure how far it departs from these principled rules, and
reuse them as the backbone of \cav{}.

\section{Threat Model}
\label{sec:threat}

\paragraph{System model}
A flow $x$ with true family $y \in \mathcal{Y}$ is analysed by $D$ detectors. Detector $d$ has
output space $\mathcal{O}_d$, either $\mathcal{Y}$ or $\{\textsc{Benign}, \textsc{Anomalous}\}$,
and emits an evidence tuple $e_d = (o_d, c_d, \pi_d)$: output, reported confidence, and
provenance (detector type and training data). An \emph{investigator} maps the evidence set
$E(x)$, optionally with verification context, to a verdict
$\hat{y} \in \mathcal{Y} \cup \{\abstain\}$, where \abstain{} escalates to a human analyst.

\paragraph{Corruption}
A corruption picks a subset $S$ of $k$ detectors and replaces their outputs with a common wrong
family
\begin{equation}
  \tilde{y}(x) =
  \begin{cases}
    \textsc{Benign} & \text{if } y \neq \textsc{Benign},\\
    a \sim \mathcal{U}(\mathcal{Y}\setminus\{\textsc{Benign}\}) & \text{if } y = \textsc{Benign},
  \end{cases}
\end{equation}
mapped into each detector's output space. The first case models a \emph{missed detection}, the
second a \emph{false alarm}. Because only tool outputs change, this models compromised,
misconfigured, poisoned, or drifting tools. It does not model traffic-level evasion, which perturbs
the flow itself and hence also the raw features that our independent channel reads
(Section~\ref{sec:discussion}). A shared wrong label is the hardest case for consensus. Under \emph{matched} confidence, $c_d$ is drawn from detector $d$'s confidences on
flows it classifies correctly, so the corruption is invisible in the confidence. Under
\emph{injected} confidence, $c_d$ is fixed to $c \in \{0.5, \dots, 0.99\}$. Under
\emph{persistent} scope, the same $S$ is corrupted for a block of flows, modelling a compromised or drifting tool. Under \emph{sporadic} scope, $S$ is drawn per
flow, modelling transient faults.

\paragraph{Assumptions and goals}
The adversary controls what tools report, but not the raw telemetry, provenance metadata,
or the investigator's instructions. The corruption is non-adaptive. We first assume clean labelled
reference data and then relax this by corrupting the reference labels (Section~\ref{sec:experiments}).
An investigator should not \emph{propagate} corrupted evidence. We measure the \emph{follow
rate}, the fraction of corrupted flows whose verdict equals the injected label, together with
accuracy (abstentions count as errors) and coverage.

\section{Experimental Testbed}
\label{sec:testbed}

\paragraph{Dataset}
We use the corrected release~\cite{liu2022error} of CIC-IDS2017~\cite{sharafaldin2018cicids},
which fixes labelling and feature-extraction errors of the original. We drop payload-less
\emph{attempted} flows and group labels into six families: \textsc{Benign}, \textsc{DoS},
\textsc{DDoS}, \textsc{PortScan}, \textsc{BruteForce} (FTP/SSH-Patator), and \textsc{Bot}.
Heartbleed, Infiltration, and web attacks (104 non-attempted flows) are too rare for per-family
evaluation. After removing identifier and time columns, invalid values, and constant columns,
and de-duplicating \emph{before} splitting, 1,790,909 flows with 82 features remain. We split them
60/20/20 with stratification and cap training data at 300,000 flows per family. Validation data serve calibration, reliability
estimation, thresholds, and prompt piloting; all results use the test split. We use a random split: CIC-IDS2017 runs each attack family on a single
day, so a split by day removes whole families, and a chronological split within a family, common in fraud
detection~\cite{yan2026temporal,wu2026classweighting}, tests
little drift, since each attack lasts 20 minutes to 7 hours and comes from one tool.

\paragraph{Detectors}
D1 is a random forest~\cite{breiman2001rf} (100 trees) and D2 XGBoost~\cite{chen2016xgboost}
(300 rounds, depth 8), both multi-class. D3 is an Isolation Forest~\cite{liu2008iforest} trained
on benign flows only, with outputs \textsc{Benign}/\textsc{Anomalous}. D4 is a rule-based IDS: a
depth-4 decision tree (at least 200 flows per leaf) whose leaves read as signatures, standing in
for signature engines that would require raw packet captures. All probabilities are calibrated
by temperature scaling~\cite{guo2017calibration}. Test accuracies are .99997 (D1), .99998 (D2),
.813 (D3), and .993 (D4). The family-level detectors disagree on 0.67\% of test flows.

\paragraph{Evidence and sampling}
Each condition corrupts a fixed, family-balanced batch of up to 5,000 test flows, which
batch-level estimators use. Statistical aggregators and \cav{} are evaluated on 420 flows per
condition (70 per family). The unverified LLM investigators were run on the first 35 flows per
family (210 flows), a subset of the 420. All comparisons use paired exact McNemar tests on shared
flows; we report 95\% bootstrap intervals.

\paragraph{LLM investigators}
All LLMs run locally through Ollama with 4-bit weights. Main models are Qwen2.5-7B-Instruct~\cite{qwen2025qwen25} and
Llama-3.1-8B-Instruct~\cite{dubey2024llama3}, plus Qwen2.5-3B and -14B for the scale analysis.
Decoding uses temperature 0, a fixed seed, and a JSON schema restricting the verdict to
$\mathcal{Y} \cup \{\abstain\}$. \textbf{Naive} receives the evidence table (tool, provenance,
output, reported confidence) and a family glossary. \textbf{CoT} additionally reasons step by
step~\cite{wei2022cot}. \textbf{\cav{}-LLM} receives the verification context of
Section~\ref{sec:method}.

\section{Conflict-Aware Evidence Verification}
\label{sec:method}

% Figure 1: CAV pipeline. Requires \usetikzlibrary{arrows.meta,positioning,shapes.geometric}
\begin{figure*}[t]
\centering
\resizebox{\textwidth}{!}{%
\begin{tikzpicture}[
    font=\scriptsize,
    node distance=4mm and 5mm,
    box/.style={draw, rounded corners=2pt, align=center, minimum height=8mm, inner sep=3pt, fill=white},
    det/.style={box, minimum width=15mm, minimum height=5mm},
    tool/.style={box, fill=blue!6},
    indep/.style={box, fill=green!8},
    llm/.style={box, fill=orange!12},
    outbox/.style={box, fill=gray!10},
    gate/.style={draw, diamond, aspect=1.6, align=center, inner sep=1pt, fill=yellow!15},
    arr/.style={-{Latex[length=1.6mm]}, thick},
    dasharr/.style={-{Latex[length=1.6mm]}, thick, dashed, green!45!black},
]
  % Input and detectors
  \node[box] (flow) {Network\\flow $x$};
  \node[det, right=6mm of flow, yshift=9mm] (d1) {D1 RF};
  \node[det, below=1mm of d1] (d2) {D2 XGBoost};
  \node[det, below=1mm of d2] (d3) {D3 IForest};
  \node[det, below=1mm of d3] (d4) {D4 Rules};

  % Evidence and the corruption that acts on it
  \node[tool, right=6mm of d2, yshift=-3mm, text width=19mm] (ev)
       {Evidence table\\$e_d=(o_d, c_d, \pi_d)$};
  \node[draw=red!70!black, dashed, rounded corners=2pt, text=red!70!black, align=center,
        above=5mm of ev, text width=21mm, inner sep=2pt] (threat)
       {Threat model: corrupt $(o_d, c_d)$ of $k$ tools};
  \draw[arr, red!70!black] (threat) -- (ev);

  % Reliability and conflict
  \node[tool, right=of ev, text width=22mm] (rel)
       {Equal-weight vote\\$\Rightarrow \bar{y}(x)$, candidates\\(confidences ignored)};
  \node[tool, right=of rel, text width=15mm] (conf)
       {Conflict\\$C(x)=1-\max_y s(y)$};

  % Independent channel
  \node[indep, below=9mm of rel, text width=40mm] (trust)
       {Trust score $T(x,y)$ on raw flow features\\(independent of all tools)};

  % Gate and outcomes
  \node[gate, right=6mm of conf] (gate) {Escalate?};
  \node[outbox, right=7mm of gate, yshift=8mm, text width=22mm] (fast)
       {Verdict $\bar{y}(x)$\\(no LLM call)};
  \node[llm, right=7mm of gate, yshift=-9mm, text width=24mm] (verify)
       {Verify: \cav{}-rule or \cav{}-LLM\\(trust, measured reliability)\\$\Rightarrow$ verdict or \abstain{}};

  % Arrows
  \foreach \d in {d1,d2,d3,d4} { \draw[arr] (flow.east) -- (\d.west); \draw[arr] (\d.east) -- (ev.west); }
  \draw[arr] (ev) -- (rel);
  \draw[arr] (rel) -- (conf);
  \draw[arr] (conf) -- (gate);
  \draw[dasharr] (flow.south) |- (trust.west);
  \draw[dasharr] (trust.east) -| (gate.south);
  \draw[dasharr] (trust.east) -| (verify.south);
  \draw[arr] (gate.north) |- node[pos=0.25, left] {no} (fast.west);
  \draw[arr] (gate.east) -- ++(3mm,0) |- node[pos=0.75, above] {yes} (verify.west);
\end{tikzpicture}%
}
\caption{Conflict-Aware Evidence Verification. Lean \cav{} ignores reported confidences, derives
candidates and conflict $C(x)$ from an equal-weight vote, and escalates a flow if $C(x) > \tau$ or
the independent trust score (green), computed from raw features the tools cannot influence,
disagrees with the vote. \cav{}-rule or \cav{}-LLM resolves escalated flows and may abstain.}
\label{fig:pipeline}
\end{figure*}
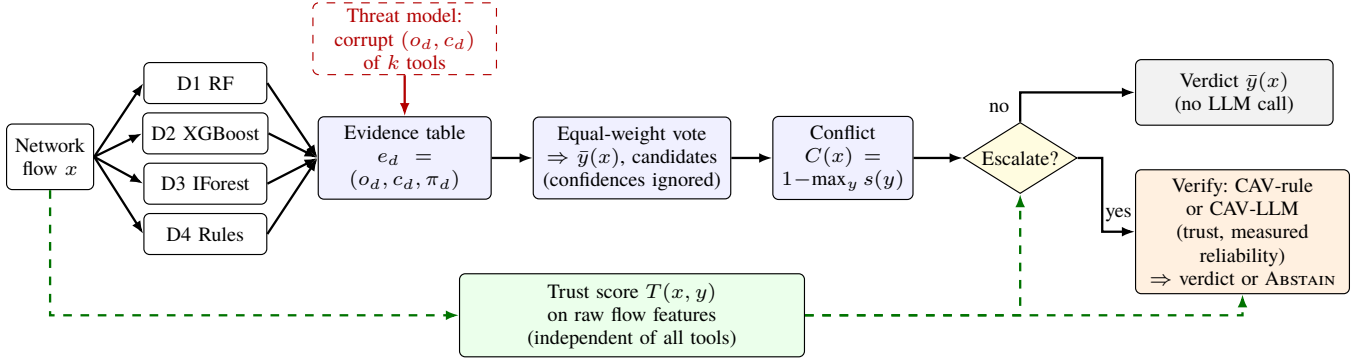

\cav{} sits between the detectors and the investigator (Fig.~\ref{fig:pipeline}). Every step is
an established technique. The contribution is their composition: reported confidences are
ignored, conflict is quantified explicitly, and only conflicting flows are verified, against
evidence the tools cannot influence. We present the lean variant used as our main system. The
submitted version (\emph{full} \cav{}) also weighted tools by estimated reliability and showed
SHAP features; Section~\ref{sec:experiments} shows that both additions hurt.

\paragraph{Candidates and conflict}
Let $\kappa_d(y)=1$ if output $o_d$ is consistent with family $y$. For the binary detector,
\textsc{Anomalous} is consistent with every attack family. The support and conflict are
\begin{equation}
  s(y) = \frac{1}{D}\sum_d \kappa_d(y), \qquad C(x) = 1 - \max_{y} s(y),
  \label{eq:support}
\end{equation}
i.e., the variation ratio. The aggregate verdict $\bar{y}(x)$ is the plurality vote, with the
binary vote split over attack families. Reported confidences enter neither, because a corrupted
tool controls them.

\paragraph{Measured reliability}
For each detector we estimate a confusion matrix $M_d[k,o] = P(o_d{=}o \mid y{=}k)$ on the
validation split. Because a tool may be compromised later, we re-estimate $M_d$ on each unlabelled
analysis window with MAP Dawid--Skene EM~\cite{dawid1979maximum}, using a Dirichlet prior centred
on the validation estimate (50 pseudo-flows per class). Lean \cav{} only \emph{reports} these
estimates to the LLM. Full \cav{} also used them in~\eqref{eq:support}, with log-odds weights
$w_d = \operatorname{clip}(\log\frac{\mathrm{acc}_d}{1-\mathrm{acc}_d}, 0, 8)$, which are optimal
for independent voters~\cite{nitzan1982optimal}, and took $\bar{y}(x)$ from the EM posterior. EM
cannot identify the honest side when at least half of the tools are corrupted in the same way, and
our analysis windows coincide with the corruption blocks, the most favourable setting for it.

\paragraph{Independent check}
\label{sec:verify}
All of the above depends on tool outputs, so \cav{} adds the trust score~\cite{jiang2018trust},
computed from raw features that the corruption does not touch:
\begin{equation}
  T(x, y) = \frac{\min_{y' \neq y} \operatorname{dist}(x, \mathcal{X}_{y'})}
                 {\operatorname{dist}(x, \mathcal{X}_{y})},
\end{equation}
where $\mathcal{X}_y$ holds up to 20,000 training flows of family $y$ (features log-scaled and
standardised). $T(x,y)>1$ means $x$ lies closer to known flows of $y$ than to any other family.
$T$ does not vote; it \emph{tests} the tools' candidates.

\paragraph{Gate and decision}
A flow is escalated if $C(x) > \tau$, if $\arg\max_y T(x,y) \neq \bar{y}(x)$, or if
$T(x,\bar{y}(x)) < 1$. Otherwise it receives $\bar{y}(x)$ without an LLM call. We use
$\tau = 0.25$, fixed on validation data before testing, so that a single dissenting tool does not
escalate a flow. On clean validation traffic this escalates 0.4\% of flows. On the test
conditions, \cav{}-rule's decisions are identical for all $\tau \in \{0, \dots, 0.5\}$, so $\tau$
only affects cost. Escalated flows are resolved in one of two ways:
\begin{itemize}
  \item \textbf{\cav{}-rule} returns whichever of the two most probable families has the higher
        trust score, if that score exceeds 1, and \abstain{} otherwise.
  \item \textbf{\cav{}-LLM} receives, per detector, the output, the reported confidence, its
        validated precision, and its re-estimated accuracy. It also receives $C(x)$ and the trust
        scores of the candidates stated in plain language. Full \cav{} adds the top-5 SHAP
        features~\cite{lundberg2017shap} of D2 for the two leading candidates. The instructions
        (prompt v2, frozen before testing) state that tool agreement is not proof, that reported
        confidence is not reliability, and that the nearest-neighbour check is independent of all
        tools.
\end{itemize}
The context deliberately omits the posterior $P(y \mid E(x))$. In a validation pilot, showing it
made the agent follow a corrupted majority on nearly every flow with three corrupted tools.

\paragraph{Post-hoc prompt revision (v3)}
The failure analysis (Section~\ref{sec:analysis}) showed that full \cav{}-LLM with prompt v2
often misread the trust scores. We revised the prompt once. It was motivated by test-set failures but
designed on validation data only, so we report it separately and base no main claim on it. The check is stated categorically
(``RESULT: supports DoS''; each candidate SUPPORTED or NOT SUPPORTED), and the JSON schema requires
the model to restate the supported family before its verdict.

\paragraph{Baselines}
We compare against majority vote (the binary vote split over attack families, ties
broken at random), confidence-weighted vote, which trusts reported confidences,
reliability-weighted (naive-Bayes) vote under $M^{\mathrm{val}}$, and Dawid--Skene alone. To test
whether \cav{} adds anything beyond access to the independent channel, \emph{+ kNN} adds the
nearest training family $\arg\max_y T(x,y)$ to the reliability-weighted vote as a fifth,
uncorrupted voter.

\section{Experiments}
\label{sec:experiments}

The main experiment crosses $k \in \{0,1,2,3\}$ corrupted detectors with both scopes under matched
confidence (Table~\ref{tab:main}); we report every condition separately, since aggregate scores can
hide regime-dependent failures~\cite{wang2026tsfm} and violations of the task's logic~\cite{wang2026choice}. Across all experiments, LLM investigators produced 33,652 test verdicts, all of which
parsed under the JSON schema.

\begin{table*}[t]
\centering
\caption{Accuracy (follow rate in parentheses); $k$ of 4 detectors corrupted, matched confidence; abstentions count as errors; best per column in bold. Statistical rows and lean \cav{} (main system): 420 flows per condition (95\% CI half-width $\le$ .05); Naive, CoT, full \cav{} (submitted version), and post-hoc v3: 210-flow subset ($\le$ .07).}
\label{tab:main}
\small
\setlength{\tabcolsep}{5pt}
\begin{tabular}{l c cc cc cc}
\toprule
 & $k{=}0$ & \multicolumn{2}{c}{$k{=}1$} & \multicolumn{2}{c}{$k{=}2$} & \multicolumn{2}{c}{$k{=}3$} \\
\cmidrule(lr){3-4}\cmidrule(lr){5-6}\cmidrule(lr){7-8}
Investigator & clean & pers. & spor. & pers. & spor. & pers. & spor. \\
\midrule
Majority vote & \textbf{1.00} & .71 (.29) & .72 (.28) & .31 (.69) & .28 (.72) & .00 (1.00) & .00 (1.00) \\
Confidence-weighted vote & \textbf{1.00} & .99 (.01) & .99 (.01) & .45 (.55) & .46 (.54) & .00 (1.00) & .00 (1.00) \\
Reliability-weighted vote & \textbf{1.00} & .98 (.02) & .98 (.02) & .53 (.47) & .55 (.45) & .07 (.93) & .07 (.93) \\
Dawid--Skene & \textbf{1.00} & \textbf{1.00} (.00) & .87 (.13) & .45 (.55) & .84 (.16) & .00 (1.00) & .00 (1.00) \\
\midrule
Reliability-weighted + kNN & \textbf{1.00} & \textbf{1.00} (.00) & \textbf{1.00} (.00) & .86 (.10) & .87 (.09) & .53 (.41) & .54 (.42) \\
kNN alone (reference) & \textbf{1.00} & \textbf{1.00} (.00) & \textbf{1.00} (.00) & \textbf{1.00} (.00) & \textbf{1.00} (.00) & \textbf{1.00} (.00) & \textbf{1.00} (.00) \\
\midrule
Naive (Qwen-7B) & \textbf{1.00} & .72 (.02) & .64 (.02) & .29 (.19) & .34 (.17) & .07 (.49) & .08 (.52) \\
Naive (Llama-8B) & .97 & .50 (.00) & .54 (.00) & .16 (.07) & .21 (.10) & .00 (.39) & .00 (.40) \\
CoT (Qwen-7B) & \textbf{1.00} & .69 (.05) & .69 (.11) & .32 (.30) & .39 (.24) & .13 (.56) & .10 (.61) \\
CoT (Llama-8B) & .99 & .88 (.01) & .86 (.00) & .35 (.19) & .42 (.17) & .02 (.50) & .02 (.56) \\
\midrule
Lean \cav{}-rule & \textbf{1.00} & \textbf{1.00} (.00) & \textbf{1.00} (.00) & \textbf{1.00} (.00) & \textbf{1.00} (.00) & .83 (.00) & .80 (.00) \\
Lean \cav{}-LLM (Qwen-7B) & \textbf{1.00} & \textbf{1.00} (.00) & \textbf{1.00} (.00) & .91 (.00) & .92 (.00) & .87 (.00) & .88 (.00) \\
Lean \cav{}-LLM (Llama-8B) & \textbf{1.00} & \textbf{1.00} (.00) & \textbf{1.00} (.00) & .90 (.04) & .89 (.04) & .84 (.08) & .84 (.06) \\
\midrule
Full \cav{}-rule & \textbf{1.00} & \textbf{1.00} (.00) & .94 (.00) & .77 (.00) & \textbf{1.00} (.00) & .45 (.00) & .82 (.00) \\
Full \cav{}-LLM (Qwen-7B) & \textbf{1.00} & \textbf{1.00} (.00) & .91 (.02) & .77 (.21) & .87 (.01) & .55 (.43) & .78 (.12) \\
Full \cav{}-LLM (Llama-8B) & \textbf{1.00} & \textbf{1.00} (.00) & .88 (.04) & .85 (.10) & .91 (.07) & .72 (.25) & .74 (.19) \\
\midrule
Full \cav{}-LLM v3 (Qwen-7B) & \textbf{1.00} & \textbf{1.00} (.00) & \textbf{1.00} (.00) & \textbf{1.00} (.00) & \textbf{1.00} (.00) & \textbf{1.00} (.00) & \textbf{1.00} (.00) \\
Full \cav{}-LLM v3 (Llama-8B) & \textbf{1.00} & \textbf{1.00} (.00) & .96 (.00) & .94 (.00) & .96 (.01) & .95 (.00) & .99 (.00) \\
\bottomrule
\end{tabular}
\end{table*}

\subsection{RQ1: Robustness of LLM Investigators}

Unverified LLM investigators are fragile. One corrupted detector reduces Naive accuracy to
.64--.72 (Qwen-7B) and .50--.54 (Llama-8B). With three corrupted detectors, every unverified LLM
falls to .13 or below. CoT helps Llama-8B at $k{=}1$ (.86--.88) but \emph{raises} the follow rate at $k{=}3$ (.50--.61 versus .39--.52). At $k \le 2$, unverified LLMs
mostly abstain rather than follow (follow rate $\le$ .19; Naive coverage at $k{=}2$ .23--.50); at
$k{=}3$, about half of the corrupted flows receive the corrupted label.

Corruption is not even needed. On 300 test flows where the \emph{honest} detectors disagree,
reliability-weighted voting, Dawid--Skene, and \cav{}-rule are at least .99 accurate. Naive Llama-8B,
however, abstains on 63\% of these flows (accuracy .37), and CoT lowers its accuracy to .23.
At the 0.67\% disagreement rate, a naive LLM investigator would escalate about 2,400 easily
resolved flows per day of this traffic~\cite{alahmadi2022false}.

\subsection{RQ2: Influence of Reported Confidence}

\begin{figure}[t]
\centering
\includegraphics[width=\columnwidth]{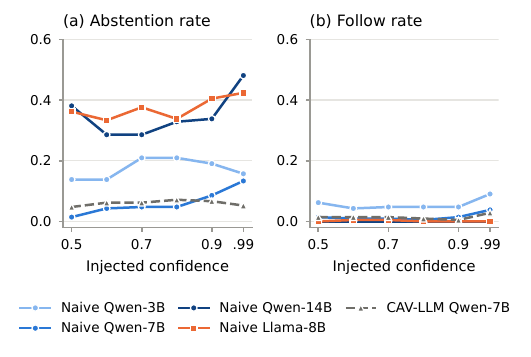}
\caption{Confidence injection ($k{=}1$, sporadic). (a) Abstention rises with the confidence of
the wrong detector; larger models do not abstain less. (b) No investigator adopts the wrong label
at a meaningful rate.}
\label{fig:confidence}
\end{figure}

We fix one corrupted detector ($k{=}1$, sporadic) and vary only its reported confidence
(Fig.~\ref{fig:confidence}). No model adopts the wrong label more often as confidence rises
(follow rate $\le$ .09). Confidence instead drives
\emph{abstention}. Qwen-7B abstains on 1\% of flows at 0.5, on 13\% at 0.99, and on 25--34\%
under matched confidence, where corrupted outputs carry near-certain confidences. Scale does not
help: Qwen-14B abstains more than Qwen-7B at every level (29--48\%). A lone confidently wrong tool thus \emph{paralyses} rather than persuades these
investigators. \cav{}-LLM is insensitive to the injected confidence (abstention 5--7\%, follow
rate $\le$ .03).

\subsection{RQ3: Conflict-Aware Verification}

Lean \cav{}-LLM, with the prompt frozen before testing, improves over the Naive investigator of the
same model in every corrupted condition (exact McNemar on the 210 shared flows, $p < 0.001$ for
all 12 pairs). With three corrupted detectors it reaches .87--.88 (Qwen-7B) and .84 (Llama-8B),
against at most .08 for Naive, and it adopts the injected label on at most .005 (Qwen) and .08
(Llama) of corrupted flows. Lean \cav{}-rule never adopts a corrupted label and is perfect up to
$k{=}2$. At $k{=}3$ it abstains when the trust score supports neither leading candidate, and its
accuracy falls to .83/.80. The LLM is significantly \emph{less} accurate than the rule at $k{=}2$
($-$.08 to $-$.11, $p < 0.001$), where it abstains on 5--8\% of flows. Only at $k{=}3$ does it help:
Qwen gains $+$.04 ($p{=}0.13$, persistent) and $+$.08 ($p{=}0.004$, sporadic), while Llama's gains
($+$.01, $+$.04) are not significant. With a perfect independent channel, the LLM therefore adds
value over a fixed rule only when most tools are corrupted.

\paragraph{Lean versus full}
Removing reliability weighting and SHAP features helps wherever two or more tools are corrupted.
On the 210 shared flows the rule gains .22 at $k{=}2$ and .37 at $k{=}3$ persistent, and the LLM
gains .06--.32 in 7 of the 8 model--condition pairs with $k \ge 2$ ($p < 0.05$). On validation data, the lean rule
is at least as accurate as the full rule in six of seven conditions, so the choice does not rest on
test data alone. An ablation of
full \cav{} at $k{=}2$ persistent locates the gain: removing reliability weights raises the rule
from .77 to 1.00, because weighting prunes the true family from the candidates
(Section~\ref{sec:analysis}); removing SHAP raises the LLM from .77 to .96; removing the trust
score drops both to .42. The lean LLM is not better than removing SHAP alone (.90/.93 versus .96/.94), so the rule drives the lean choice.

\paragraph{Verification versus an extra voter}
The nearest-neighbour signal as a fifth voter makes reliability-weighted voting much stronger
(.86--.87 at $k{=}2$, .53--.54 at $k{=}3$), so the independent channel is the key ingredient. With
most tools corrupted, however, the channel is outvoted (follow rate .41--.42). Lean \cav{}-rule beats it by .13--.14 at $k{=}2$ and
.27--.30 at $k{=}3$, and lean \cav{}-LLM by .30--.35 at $k{=}3$ (all $p < 0.001$). Using the
independent channel to \emph{test} candidates, rather than to outvote the tools, keeps the
corruption out of the verdict.

\begin{figure}[t]
\centering
\includegraphics[width=\columnwidth]{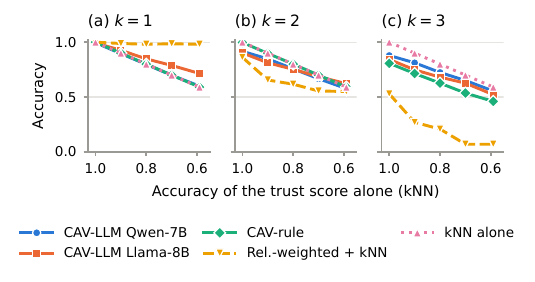}
\caption{Lean \cav{} with a noisy trust score (5--30\% of reference labels corrupted), $k$
corrupted tools, 420 flows per point (both scopes).}
\label{fig:degraded}
\end{figure}

\paragraph{An imperfect independent channel}
We move 5--30\% of the trust score's reference labels to a random other family and rerun lean
\cav{} (210 flows per condition), which lowers the accuracy of kNN alone to .90--.59
(Fig.~\ref{fig:degraded}). With at most one corrupted tool, \cav{} inherits the channel's errors:
its accuracy tracks kNN alone, even at $k{=}0$, where four correct tools are overruled, whereas the
+kNN vote stays at $\ge$ .98 because the tools outvote the channel. On flows where kNN
is wrong, the rule abstains on 87--92\%, Qwen adopts the wrong check (accuracy $\le$ .04), and
Llama recovers only .14--.17, whereas the +kNN vote recovers about half.
The ordering reverses with more corruption. At $k{=}2$ \cav{} is ahead at every noise level, and at
$k{=}3$ \cav{}-LLM reaches .81--.52 against .27--.07 for the +kNN vote. The LLM's advantage over
the rule at $k{=}3$ grows to $+$.10--.11 for Qwen ($p < 0.001$) and $+$.03--.09 for Llama
($p < 0.05$ from 10\% noise). Verification keeps corrupted
\emph{tools} out of the verdict, but not errors of the channel: it moves trust to the check.

\paragraph{Post-hoc prompt (v3)}
Applied to full \cav{}-LLM, the revised prompt reaches 1.00 in every condition for Qwen-7B and
.94--1.00 for Llama-8B. A corrupted label is followed on only 2 flows, both by Llama-8B at
$k{=}2$ sporadic (the .01 in Table~\ref{tab:main}). In both, the model restated the correct family
and then overrode it because two tools agreed with high confidence. Since kNN alone is also
perfect here and v3 was motivated by test-set failures, we treat it as a hypothesis, not a result.

\subsection{RQ4: Cost}

Unverified LLMs call the model on every flow. Like cloud--edge agents that hand off to a larger
model only above an uncertainty threshold~\cite{huang2026agent}, lean \cav{} calls it only on escalated flows: 0.7\% of clean test flows, 54\% of flows with one corrupted detector, and every flow with two or more.
Calls are longer (about 580 versus 300 prompt tokens; 3.4 versus 2.3\,s on a laptop).
Full \cav{} escalated far less (2\% at $k{=}1$ persistent) because reliability weights silenced
the dissenting tool, the same mechanism that caused its errors. Escalation also exposes the LLM's habit of abstaining on honest disagreement.
Lean \cav{} escalates 158 of the 300 naturally disagreeing flows, mostly two-against-two splits
involving the weak Isolation Forest. Qwen resolves them (.99), but Llama abstains on 85 (accuracy
.71), calling the check ``inconclusive''. \cav{}-rule resolves them all at .99. On mostly clean traffic, \cav{} needs about two orders of
magnitude fewer LLM calls than an always-on investigator.

\section{Failure Analysis}
\label{sec:analysis}

\paragraph{Unverified LLMs count tools}
Accuracy alone does not show how a verdict was reached~\cite{sun2026beyond}, so we read the rationales.
When unverified LLMs abstain on corrupted evidence, 94\% of their rationales cite conflict between
tools, even when three of four agree; when they follow a corruption, 95\% cite the number of
agreeing tools. Their verdict is a function of how many tools agree, which is exactly what the
corruption controls.

\paragraph{\cav{}-LLM misreads rather than misses evidence}
In all 307 escalated flows where full \cav{}-LLM adopted the corrupted label, the trust score
pointed to the true family. Yet 88\% of these rationales invoke the nearest-neighbour check in
support of the \emph{wrong} label (``nearest-neighbour check supports Benign over DDoS'' for a DDoS
flow whose trust score favoured DDoS). Lean \cav{}-LLM follows far less often (100 flows, 94 by Llama-8B)
with the same pattern: the trust score was right in all, and 91\% of rationales cite it. \cav{}-rule,
reading the same number, never follows.

\paragraph{More evidence can hurt}
The aggregate posterior is derived from corrupted outputs but looks like a single confident
summary, and the LLM defers to it. SHAP features rarely discriminate between two attack
candidates, since every attack deviates from the benign median, and they distract from the trust
score; prototype-based text classifiers likewise find explanations by similarity to known training
cases easier to use than feature attributions~\cite{wen2026transformer,wen2025gaprotonet}. Reliability weighting backfires for a related reason. With two tools persistently
corrupted alike, Dawid--Skene assigns the corrupted side near-certain reliability: only 58\% of
flows are escalated, and the true family is among the two verification candidates for only 60\% of
them. With equal weights, every flow is escalated and the truth is a candidate in 99.5\% of cases. As in retrieval fusion, adaptive weights need not beat a fixed
blend~\cite{wu2026evidenceunit}. Verification can only confirm candidates that the upstream
aggregation keeps alive, as early hard selection likewise propagates errors in retrieve-then-plan
pipelines~\cite{ye2026osef}, so candidate generation should be robust rather than optimal for
honest tools. This is why lean \cav{} drops both components.

\section{Discussion and Limitations}
\label{sec:discussion}

\paragraph{What the contribution is}
Our primary contribution is the measurement in RQ1--RQ2 and the design rules it yields, not a
claim that an LLM out-predicts its evidence; a fixed rule is the most accurate investigator in
most of our conditions. The LLM helps only where that rule abstains:
when most tools are corrupted. \cav{} is also only as good as its channel: with a noisy channel and few corrupted tools, adding
the channel as a voter is better (Fig.~\ref{fig:degraded}).

\paragraph{Implications}
Four design rules follow.
\begin{itemize}
  \item Treat a tool's reported confidence as an untrusted field, and do not let estimated
        reliability prune the candidates that verification will test.
  \item Withhold derived summaries, which look like independent evidence, and state independent
        evidence as a categorical result.
  \item Keep an evidence channel the tools cannot touch (raw flows, packet samples, endpoint
        telemetry) and measure its accuracy: consensus fails when enough tools are corrupted
        alike, and verification is only as good as the channel.
  \item Spend LLM reasoning only where tools conflict, and prefer a fixed, auditable rule where it
        suffices, as regulated fraud modelling prefers auditor-readable encodings that cost little
        accuracy~\cite{han2026encoders}.
\end{itemize}
The same structure arises in anti-money-laundering triage, fraud review~\cite{yan2026target}, and threat-intelligence
enrichment.

\paragraph{Limitations}
CIC-IDS2017 is a 2017 laboratory dataset, and our random split overstates deployed detector
accuracy~\cite{arp2022dos}. Three rare families are excluded, and we do not evaluate unseen
families. For a novel attack, the trust score can at best support no known family, which \cav{}-rule turns into an abstention. Corruptions are synthetic and non-adaptive. Traffic-level evasion~\cite{apruzzese2022realistic} perturbs the raw features that D1, D2, D4, and
the trust score share, so our channel is independent of tool compromise but not of evasion; that
requires telemetry the attacker cannot shape. Random reference-label noise is only a first step towards targeted poisoning.
We evaluate open-weight models up to 14B
with 4-bit quantisation; larger models may weigh evidence differently, and
trust-related scores drift even across releases of one model line~\cite{fan2026movingtarget}.

\section{Conclusion}
\label{sec:conclusion}

Open-weight LLM investigators degraded sharply once one of four detectors was corrupted. \cav{} recovers most of the lost accuracy at a fraction of the
LLM cost by testing the tools' claims against evidence they cannot influence. The cost is a new
point of trust: with a noisy independent channel, \cav{} inherits its errors, so that channel must
be measured in turn.

% Required by the ACSAC 2026 AI policy; does not count towards the page limit.
% AUTHORS: review and edit this statement so that it accurately describes your own use.
\section*{LLM Usage Statement}

\emph{LLMs as research subject.} The investigators evaluated in this paper are open-weight LLMs
(Qwen2.5 3B/7B/14B, Llama-3.1-8B) run locally. Their weights are public, decoding is
deterministic, and every response is cached, so all reported results can be regenerated exactly.

\emph{LLMs as a writing and engineering aid.} The authors used a proprietary LLM-based coding
assistant (Claude, Anthropic) to help implement the experimental pipeline, run and analyse
experiments, draft and edit text, and check bibliographic entries against publisher records. All
experimental results were produced by the authors' code from stored predictions, and every
number in the paper was checked against those outputs. The authors reviewed all text, code, and
references and take full responsibility for the content. Because the assistant is not openly
available, its contributions cannot themselves be reproduced. Our code, configuration,
and response cache make the reported results reproducible independently of the assistant.

\bibliographystyle{IEEEtran}
\bibliography{refs}

\end{document}